\documentclass[namedreferences,hyperref,optionalrh]{spr-sola}
\usepackage{longtable}
\usepackage{graphicx}    
\usepackage{rotating}
\usepackage{color}
\usepackage{multicol}

\usepackage{xcolor}

\chardef\us=`\_

\begin{document}

\begin{frontmatter}
\title{Origin of Joy's Law in the context of Near-Surface Convection on the Sun}

\author[addressref=aff1,email={hannah.schunker@newcastle.edu.au}]{\inits{H.}\fnm{Hannah}~\snm{Schunker}\orcid{0000-0001-9932-9559}}
\author[addressref=aff1]{\inits{A.L.}\fnm{Asha Lakshmi}~\snm{K~V}\orcid{0009-0000-8340-6505}}
\address[id=aff1]{The University of Newcastle, Australia}

\runningauthor{Schunker and K V}
\runningtitle{Joy's Law in the context of Convective Flows on the Sun}

\begin{abstract}
Joy's law is a well-established statistical property of solar active regions that any theory of active region emergence must explain.
This tilt angle of the active region away from an east-west alignment is a critical component for converting the toroidal magnetic field to poloidal magnetic field in some leading dynamo theories, and  observations show they are important for the reversal of the sign of the global solar magnetic dipole.
This review aims to synthesise observational results related to the onset of Joy's law, placing them within the broader context that describes active region emergence as a largely passive process occurring near the surface of the Sun.
\end{abstract}
\end{frontmatter}

\section{Introduction}
     \label{sec:introduction} 

It has become increasingly clear that convective flows play an important role in the emergence of magnetically active regions on the Sun \citep[see][and references therein]{Birchetal2016,Weberetal2023}.
The generation of active region tilt angles is a central aspect of these models and warrants re-examination. In this paper we will review the compatibility of  this framework with the characteristics of Joy's law.

Joy's law is the statistical trend where the leading polarity (relative to the direction of the Sun's rotation) of an active region tends to be closer to the equator than the following polarity \citep{Hale1919}, and the tilt angle away from an East-West orientation tends to increase with the sine of the (unsigned) latitude. Traditionally, Joy's law has been measured from intensity continuum observations of sunspots.  \cite{Leighton1969} parametrised Joy's law as $\sin \gamma = 0.5\sin \lambda$ (where $\gamma$ is the tilt angle and $\lambda$ is the unsigned latitude) and other closely related functional forms have also been measured  \citep[e.g.][]{McClintockNorton2013, Pevtsovetal2014,Wang+2015}. 
Statistically, the quantitative determination of Joy's law from magnetic field maps yields a similar dependency of tilt angle on latitude to that determined from continuum white-light observations \cite[e.g. $ \sin \gamma = 0.48 \sin \lambda + 0.03$,][]{WangSheeley1991}, with the advantage of avoiding any misidentification of polarity pairs \citep[see also][]{Wang+2015,Will+2024}. 
Due to the sine-latitude dependence of the tilt angle, the Coriolis force is widely thought to be the underlying cause of Joy's law.
The average tilt angle of active regions over the whole solar cycle (and thus over all latitudes) was found to be anti-correlated with the amplitude of the cycle \citep{DasiEspuigetal2010}, suggesting that the tilt angle may play an important role in setting the amplitude of the solar cycle.

\subsection{Importance of Joy's Law for the solar cycle}\label{sec:cycle}

It is commonly accepted that active regions on the Sun form from coherent magnetic flux tubes which rise as a loop from the interior and through the surface creating the observed positive and negative polarities \citep[e.g.][]{Fan2009,Cheung2014}. The coherent magnetic flux tubes presumably originate from the Sun’s global toroidal magnetic field, which lies deeper below the surface \cite[e.g.][]{Charbonneau2020,Weberetal2023}. In dynamo theory, the $\alpha$-effect is the process which imposes a writhe on the toroidal magnetic field   \citep[e.g.][]{SteenbeckKrause1969,Cameron+2017}. In Babcock-Leighton dynamo models \citep[after][]{Babcock1961,Leighton1964}, the relevant writhe is the geometric deformation of the flux tube associated with Joy's law, which is the key process that converts the global toroidal magnetic field component to the poloidal \cite[e.g.][]{Cameronetal2015,Cameronetal2017,Biswas+2022}.
 
In turn, the polar field is the seed for the toroidal field of the next solar cycle, and thus the amplitude of the cycle is strongly correlated with the preceding polar field \citep[e.g.][]{MunozJaramillo+2013}. What sets the amount of flux at the poles of the Sun is the flux that is advected to the poles from the surface magnetic field via the meridional flow.
In determining the strength of the polar fields, the amount of cross-equatorial transport of the field at the level of the photosphere is crucial \citep{Durrant+2004}.
The evolution after the emergence process is well described by surface flux transport (SFT) models  \citep[e.g.][]{Yeates+2023}, which is a linear process. A non-linearity is required to limit run-away cycle amplitude growth, and one possibility is the source term of the SFT model, that describes the properties of the magnetic flux just after emergence.
Two possible non-linearities, `tilt quenching' and `latitude quenching' of the active regions both modulate the amount of flux that crosses the equator \citep[for a discussion of other possible non-linearities see][]{CameronSchussler2023}.

Tilt quenching modulates the amount of net flux which crosses the equator by assuming that the average tilt angle depends on the amplitude of the cycle. A strong cycle in terms of magnetic flux is assumed to lead to a smaller average tilt angle, which is less efficient at transporting net flux across the equator \citep[and observationally supported by][]{DasiEspuigetal2010}. In contrast, latitude quenching achieves much the same thing by assuming that active regions emerge on average at higher latitudes, further from the equator. Both mechanisms assume that the emergence properties change so as to reduce the amount of flux crossing the equator during strong cycles. However, the observational \citep[e.g.][ suggests the tilt angle is not dependent on flux]{Schunkeretal2020} and theoretical \citep{Talafha+2022} evidence of either mechanism is ambiguous.
Thus, understanding the physical mechanism that sets the tilt angle is essential because it determines how the solar dynamo self-regulates through non-linear feedbacks \cite[see also][]{Cameron+2013}. 



\subsection{Importance of convection for active region emergence}\label{sec:conv}
Due to computational limitations, dynamo and flux emergence models of the full spherical solar convection zone typically span only about three pressure scale heights above the base of the convection zone, extending up to about $0.97$~R$_\odot$ \citep[e.g.][]{FanFang2014}. Flux emergence simulations that include the surface typically cover the intervening pressure scale heights \citep[e.g.][]{RempelCheung2014}, although the overlap is increasing \citep[e.g.][]{HottaIijima2020}.  In this review we will restrict our focus to the near-surface, and assume that the flux tubes are formed a priori, either deeper below or within the convective simulation domain depending on the model.

 In many successful models of flux emergence, the magnetic field drives the emergence process  \cite[see][for an in-depth review]{Fan2021LRSP}. This is what we define as an \emph{active} emergence. However, \cite{Birchetal2016} demonstrated that the flows associated with active region emergence are  on the order of, or less than, the near-surface convective flow velocities, in contrast to the predictions from an active emergence process.
 A growing body of work  supports the findings that the convective flows are important for the emergence process \citep[e.g.][]{HottaIijima2020, Gottschling+2021,  Mani+2024, Schunker+2024}. 
 The current paradigm, largely driven by observational constraints, points to a more \emph{passive} process, at least near the surface.

\cite{Schunkeretal2016} found that the east-west separation of the polarities is anti-symmetric relative to the local solar surface rotation rate, indicating that the flux tubes are embedded within the bulk local plasma flows.
Active regions take, on average, about 2 days to grow to their maximum magnetic flux at the surface of the Sun and have an average east-west extent of about 40~Mm \citep[e.g.][]{Weberetal2023}. These scales are similar to supergranules, one of the dominant scales of convection on the Sun, which have a lifetime of 1-2 days and a size of 20-40~Mm \citep{RinconRieutord2018}. Indeed, \citet{Birchetal2019} found that active regions prefer to emerge associated with east-west aligned converging flow regions reminiscent of inter-supergranular lanes.
Further investigating this, \citet{Schunker+2024} confirmed that, regardless of where in the supergranulation pattern an active region emerges, statistically they are associated with an additional flow signature: about 1~day before emergence all active regions show a converging flow of about 20~m/s, an order of magnitude smaller than the supergranulation flows, followed by an increase in the diverging horizontal flows. This increase in diverging horizontal flows could indicate that active regions are brought up to the surface with upwelling supergranules. Together, these findings underscore the important role of convective flows in active region emergence. Yet, Joy's law imposes an additional constraint that has not yet been addressed in this framework.

Sunspots only form in active regions once sufficient magnetic flux has emerged $\approx 2$~days after the associated magnetic bipole first appears at the surface \citep[e.g.][]{BumbaSuda1984,Schlichenmaier+2010}, and so understanding the onset of Joy's law requires observations of active region magnetic fields before they form sunspots.
In this review, we will present the current state of understanding of the onset and evolution of Joy's law, in terms of observations and the constraints they place on models of flux emergence that attempt to explain Joy's law.

\section{Origins of Joy's Law}\label{sec:origin}

Given that both the active region tilt angles and the Coriolis force depend on the sine of the latitude, the Coriolis acceleration, $\mathbf{a_C} = -2 (\mathbf{\Omega \times v})$ (where $\mathbf{v}$ is the velocity relative to the rotating frame with angular rotation rate, $\mathbf{\Omega}$) is widely regarded as a likely driver of Joy's law.
The key question, however, is which specific flows the force acts upon to cause Joy's law:
is the dominant effect directly on the flows associated with the rising flux tube \citep[e.g.][]{DSilvaChoudhuri1993}, or on convective flows, where the tilt angle is the result of a passive reaction of the flux tube to the horizontal vorticity \citep{Parker1955,Weberetal2011}.

In the thin flux-tube model, a tilt angle is generated by the Coriolis force acting on retrograde flows  on the order of $\sim 100$~m/s within the flux-tube at the apex  of the loop as a result of angular momentum conservation \citep[e.g.][]{Abbett+2001,Fan2008,RempelCheung2014}, as they rise through the convection zone. The rise speed depends on the magnetic buoyancy of the tube, and hence on the magnetic flux. Although there were earlier indications that the tilt angle depended on the magnetic flux of the active region on the Sun \citep[e.g.][]{Fisher+1995},  statistical studies of hundreds of active regions do not show a dependence of the tilt angle on  magnetic flux  \citep[e.g.][]{StenfloKosovichev2012,Schunkeretal2020}. This suggests that Joy's law is not a direct consequence of an active process.

Another possibility to explain Joy's Law is the conservation of magnetic helicity in a flux tube as it rises through the convection zone \citep[e.g.][]{Berger84,LongcopeKlapper1997}. The magnetic helicity is composed of the writhe, which measures the deformation of the flux tube axis, and the twist of the magnetic field lines about the axis. Magnetic helicity is a conserved quantity, and  changing one component necessarily requires a change in the other. This is inherently an active process, however there are arguments that deep below the surface a flux tube gains twist through interaction with the surrounding turbulent velocities which have a non-vanishing kinetic helicity,  and subsequently develops a writhe \citep{Longcope+1998}.  
While there have been multiple efforts to constrain the helicity and twist of the surface magnetic field in active regions \citep[e.g.][and references therein]{Pevtsovetal2014}, any observed relationship between the twist and writhe (and thusly tilt) is still ambiguous, and so for the purposes of this review we will consider the model of writhe generating Joy's law an active process.

\section{Onset of Joy's Law}\label{sec:onset}

Understanding how Joy's law develops during active region emergence at the surface of the Sun is the most direct constraint on the underlying mechanism available. Capturing the  earliest stages of active region emergence and the evolution of Joy's law is only possible by observing the magnetic field at the surface of the Sun. Accordingly, the remainder of this review will focus on insights gained from surface magnetic field observations.

\citet{KosovichevStenflo2008} and \citet{Schunkeretal2020}  found that, on average, active regions emerge with east-west aligned polarities. Other studies have measured a small average tilt angle in the direction of Joy's law \citep[e.g.][]{Will+2024,Sreedevi+2024}.  Nonetheless, each of the recent large statistical studies have consistently shown that the average tilt angle increases during the emergence process.  Furthermore, \cite{Schunkeretal2020} found that the latitudinal dependence of the tilt angle is only reflected in the extent of the active region in the north-south direction. 
These findings stand in contrast to predictions of the thin flux tube model and induced writhe model, where the tilt angle is the result of a geometric phenomena set below the surface by the deformation of the flux tube. 
Thus, assuming the tilt angle is generated by the Coriolis force,  it must be acting on some east-west flows near the surface to induce a north-south separation of the active region polarities.

\cite{RolandBatty+2025} modelled the Coriolis effect in a three-dimensional Cartesian magnetohydrodynamic simulation of a flux tube ascending from a depth of 11~Mm below the surface. On the Sun, Joy's law is weak and is only evident as an average over many active regions. To achieve a measurable effect in a single simulation, they considered a rotation rate $\sim 100$~times faster than the Sun.   The flux tube  emerges at the surface with a tilt angle consistent with Joy’s law when scaled to the Sun's slower rotation, showing that it is feasible that Joy's law could be generated by the Coriolis force acting on horizontal flows in the near-surface layers.

\section{Uncertainties in measuring Joy's law}

Joy's law is only evident after averaging over a large sample of  active region tilt angles, and the measured tilt angle for any individual active region can be substantially different to the expected value from Joy's law.
This large scatter in the tilt angle is consistent with inherent buffeting of the magnetic field by the convective flows \citep[e.g.][]{Weberetal2011}. 
It is probable that the inconsistencies in results between observational studies \citep[for example][]{KosovichevStenflo2008,Schunkeretal2020,Will+2024,Sreedevi+2024}, are due to systematics within the limits of this uncertainty. 

There are numerous differences between the analysis methods, most of which we expect to have little impact on the measurement of the average tilt angle, however, the method of identifying the polarity locations, and the definition of the emergence time are important and we discuss their implications here.
When measuring the onset of the tilt angle, we aim to measure the tilt angle from the very start of the active region emergence process. At this stage, the active region bipoles are small features relative to the pixel size of the observed magnetic field map and close together, such that a small displacement in the relative location of the polarities results in a large change in tilt angle, increasing the uncertainty.

On timescales of fractions of a day, defining the emergence time is also important.
Inspired by previous helioseismic studies \citep{Lekaetal2013}, \citet{Schunkeretal2016} defined the emergence time as the time when the active region had an unsigned magnetic flux  with 10\% of the value 36~hours after being labelled as an active region with an NOAA number. This definition of the emergence time was determined to within the time cadence of 12~minutes of the observations. 

\begin{figure}[H]
    \centering
        \includegraphics[width=0.8\textwidth]{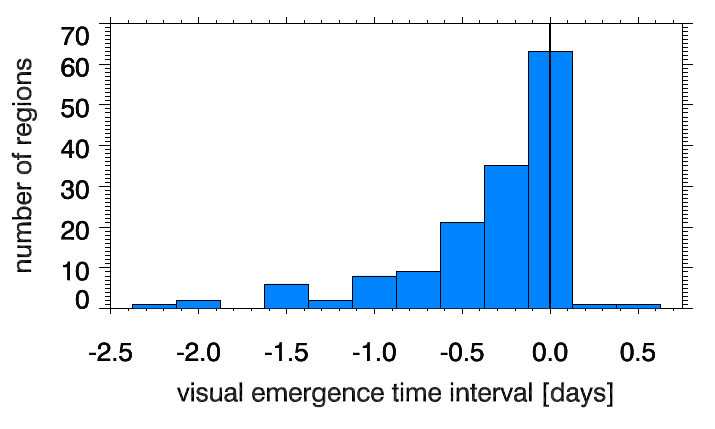}
    \caption{Number of active regions with visual emergence time interval relative to the nominal emergence time interval \citep[as defined in the SDO/HEARS;][]{Schunkeretal2016}. About half of the active regions in the sample can be visually identified as emerging up to 2 days before the nominal emergence time.}
    \label{fig:VET}
\end{figure}

Recognising that the definition of emergence time defined in the SDO/HEARS \citep{Schunkeretal2016} can be more than two days after the appearance of the magnetic bipole associated with the active region \citep[see][]{AlleySchunker2023}, we defined a visual emergence time for each active region. To do this, we used the Postel projected line-of-sight magnetic field maps averaged over $\approx 6$~hour time intervals \citep[as described in][]{Schunkeretal2016} and specified the location of the centre of the bipole at the first time interval it was visible. Figure~\ref{fig:VET} shows a histogram of the visual emergence times as a function of time interval. 

To isolate the effect of changing only the definition of the emergence time, we retained the measured emergence locations as defined in \citet{Schunkeretal2016} and so the polarities are not centred.

\begin{figure}[H]
    \centering
        \includegraphics[width=0.8\textwidth]{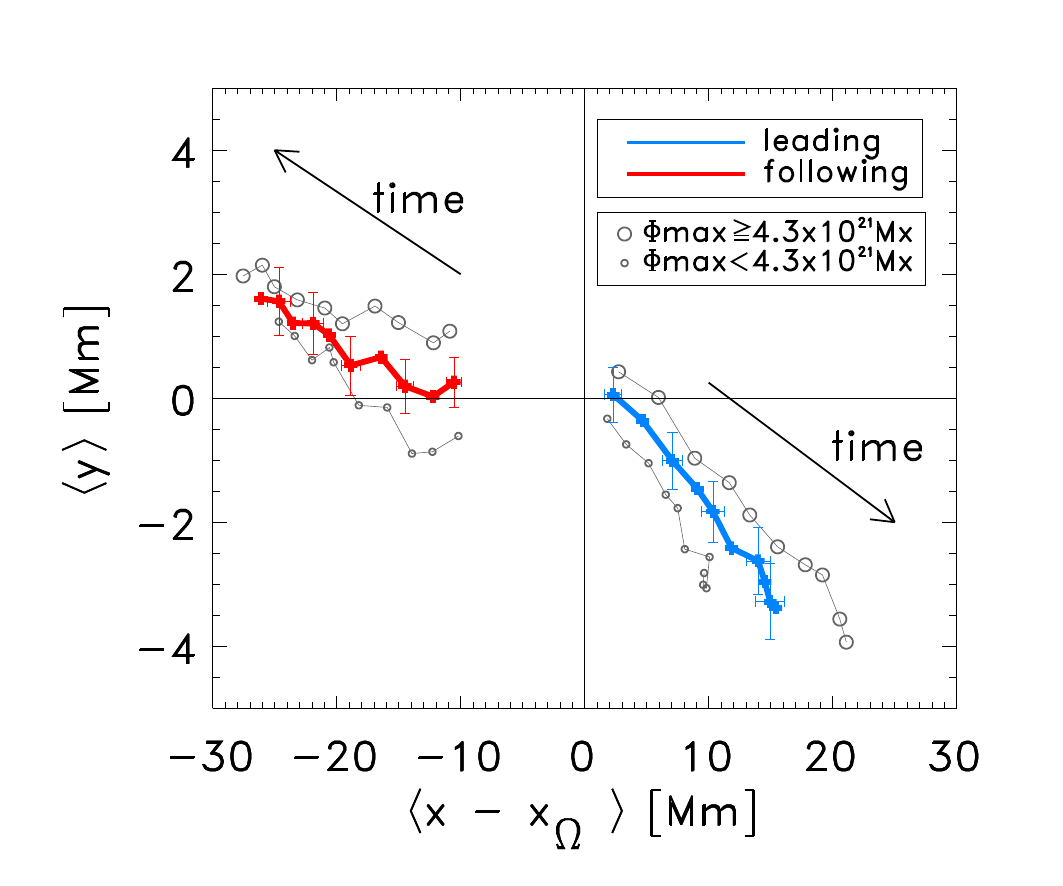}
    \caption{Average motion of the polarities during active region emergence.  The red and blue data points closest to the centre are averages of the polarity locations at the  visual emergence time (as defined in Fig.~\ref{fig:VET}), and then subsequent time intervals are further and further apart.  The tilt angle grows in time as the polarities move further apart \citep[as shown in][]{Schunkeretal2019,Schunkeretal2020}.  }
    \label{fig:pos}
\end{figure}

We found that there is a small average separation of the polarities in the north-south direction, which corresponds to  a small tilt angle in the direction of Joy's law at the beginning of the emergence (Fig.~\ref{fig:pos}). 
In agreement with Fig.~2 in \citet{Schunkeretal2019} and Fig.~A.3 in \citet{Schunkeretal2020}, the polarities emerge with an orientation consistent with an east-west alignment and the tilt angle grows as more and more magnetic flux emerges onto the surface.
At the visual emergence time, regions with a higher magnetic flux have a larger tilt than the low flux regions. This suggests a flux dependence, but again, these differences are subtle and within the noise.
We conclude that, in this case,  the small tilt angle is consistent with an east-west orientation within the uncertainties and the general statement that the tilt angle grows as active regions evolve remains true. More statistics would help to conclusively determine whether there is a flux dependence.

Joy's law is ultimately determined from the position of each polarity in an active region relative to a constant latitude.
Determining the location of the polarities using a centre of mass or centre of gravity approach, with some threshold for unsigned magnetic field strength is sensitive to the spatial mask. Transient flux appearing and disappearing,  flux moving in or out of the mask or another bipole emerging within the mask can have a substantial effect on the measured location of the polarity making fully automated algorithms challenging and in practice requires a human-in-the-loop to verify the result. \cite{Schunkeretal2020} first identified the location of the polarities about 12~hours after emergence when the active regions consist of well-formed bipoles. Once the location of the bipoles was measured at this time, then these specific features  were tracked forward and backward in time, regardless of transient magnetic flux moving within the spatial mask. This is an important point, particularly for complex active regions or `double emergences' where two strong bipoles emerge close to one another and rearrange themselves into a large complex active region, like the well-studied AR~11158.

Differences in studies may also be due to real selection bias of the active regions in statistical data sets, cadence of the observations, the spatial resolution of the instrument, the pixel size, and  method of measuring the location of the polarities.
We emphasise that  it is essential that the implications of these differences, and the definition of the emergence time and polarity location are clearly specified to allow unambiguous comparisons between independent studies.  These systematic uncertainties, however, may not be an issue if the random uncertainties introduced from buffeting by the convective flows are larger.

\section{Summary}
Joy's law must be understood in the context of the convective flows in which the flux tube is embedded \citep[e.g.][]{RolandBatty+2025} to constrain the emergence process.
The active models (i.e. thin flux tube and helicity conservation models) predict that active region tilt angles are primarily a geometric effect of the deformation of the tube set below the surface. 
The growth of the average active region tilt angle observed during the emergence process \citep{KosovichevStenflo2008,Schunkeretal2020,Will+2024,Sreedevi+2024}, however, points to a near-surface effect. 
\citet{RolandBatty+2025} demonstrates that this is feasible, invoking the Coriolis force acting on near-surface convective flows. 

If we assume that the polarities are predominantly influenced near the surface by supergranule-type east-west flows, then we can estimate the expected north-south displacement. For $v_x=-250$~m/s and a lifetime of $T=2$~days \citep{RinconRieutord2018}, at a latitude of $\lambda=20^\circ$, and a rotation rate $\Omega = 480$~nHz (a rotation period of 25~days), the expected displacement in the north-south direction would be  $\delta_y = - \Omega  v_x  \sin(\lambda)  T^2 \approx  1.5$~Mm, which is on the order of both the north-south displacement of the following and leading polarity in Fig.~\ref{fig:pos}, showing that this concept is plausible. 

Understanding the physical mechanism that governs the tilt angle is important, as it may represent the primary non-linear feedback that enables the solar dynamo to self-regulate.
The new paradigm that active region formation is a passive process, has rejuvenated an exploration of alternative origins for flux tubes in the Sun. Global convective-dynamo simulations are able to generate buoyant magnetic loops from large-scale convection \citep{FanFang2014}, or large-scale `wreaths' of flux, that have a statistical trend of tilt in the same direction as Joy's law and with comparable scatter \citep{Nelson+2014}. However, a convincing simulation of the global magnetic field that spawns active region progenitors and reproduces all of the observed features of the solar-like activity cycle remains elusive \citep{Charbonneau2020}.

 Despite the inherent limitations, models such as the thin flux tube model remain appealing for their simplicity and success in providing a physical explanation for many of the statistical properties of active regions \citep[e.g.][]{Caligari+1995,Fan2008,Weberetal2011,Fan2021LRSP}. 
To be thin, the flux tube must be thinner than the relevant length scales e.g. pressure, density, radius of curvature, but the magnetic field can vary along its length. By specifying only the flux and field strength at the base of the convection zone, these models can explain the tilt angle, the tendency for the leading polarity to form a sunspot \citep{BrayLoughhead1979}, and a radial emergence at the appropriate latitudes.  However, these simulations are not valid in the top 20~Mm beneath the surface, and \citet{RolandBatty+2025} demonstrated that a feasible alternative is that Joy’s law could be generated in the near-surface layers.

While not excluding the possibility of an $\alpha$-effect operating in the convection zone, we suggest that instead of the poloidal component of the magnetic field being generated by the presence of toroidal field, that it is instead a result of turbulent transport of toroidal field through the surface \citep[a concept also supported by][]{Cameron+2025}.

Assuming the Coriolis force is causing the tilt angle, a counteracting force due to a combination of magnetic tension, drag force and advection  must be acting to to halt the motion of the polarities, as observed \citep[see also][]{Schunkeretal2019}.
Magnetic tension is proportional to the square of the magnetic field and the curvature; the drag force is proportional to the cross-sectional area of the polarity, the square of the speed of the polarity relative to the surrounding plasma, the plasma density and Reynold’s number; and the advection is proportional to the local flows.
Understanding the motion of the polarities in relation to these quantities will help us to better constrain the origins of Joy's law.

Future numerical simulations of active region formation should aim to differentiate the relative contributions of the Coriolis effect acting on near-surface convective flows passively tilting the polarities, the Coriolis effect acting on diverging flows near the subsurface apex of magnetic flux tubes, and the coherent contributions of magnetic twist and writhe in the flux tubes.

Increased observational statistics are essential to more concretely understand any dependence on magnetic flux. Anti-Joy's law regions are an equally important sample that must also be explained by the same underlying mechanism.
A larger sample of active regions is required to reduce the random noise in order to be able to do differential statistical analysis. Given that on the order of 100-200 active regions has convincingly demonstrated the importance of the flows, a factor of ten increase would serve as a sufficient test set \citep[e.g. similar to][]{Will+2024,Sreedevi+2024}.
Since we do not know exactly where or when an active region will emerge, pursuing this problem requires high cadence, high duty cycle, high spatial resolution observations of the magnetic field at the surface of the Sun. Monitoring missions, such as the Michelson Doppler Imager onboard the Solar and Heliospheric Observatory \citep{SOHO1995} and Helioseismic and Magnetic Imager \citep{scherreretal2012,Schou+2012} onboard NASA's Solar Dynamics Observatory (SDO/HMI), are the ideal instruments to pursue this problem. While there are no designated successor missions currently proposed, SDO has the potential to remain operational well into the next solar cycle, ensuring continued valuable data of unprecedented quality, until suitable ground-based telescope networks become operational \cite[e.g.][SPRING, ngGONG]{Gosain+2018,Hill+2019} .

\begin{acks}
We acknowledge support from  the Australian Astronomical Optics Data Central (Macquarie University) which hosts the OzSDO database \url{https://datacentral.org.au}.
HS and ALKV  acknowledge the Awabakal people, the traditional custodians of the unceded land at the University of Newcastle on which their research is conducted and this review was written.
\end{acks}

\begin{fundinginformation}
HS is the recipient of an Australian Research Council Future Fellowship Award (project number FT220100330) and the research behind this review is partially funded by this grant from the Australian Government.
HS also gratefully acknowledges the support for the research contributing to this review through an Australian Research Council Discovery Project DP230101240.
\end{fundinginformation}

\begin{authorcontribution}
HS led the construction of the review, providing the foundational framework, while ALKV contributed her expertise to contextualise the definition of emergence times within the existing literature. Both authors reviewed the manuscript.
\end{authorcontribution}







 \begin{ethics}
 \begin{conflict}
 The authors declare that they have no conflicts of interest.
 \end{conflict}
 \end{ethics}


\bibliographystyle{spr-mp-sola}
\bibliography{main_HS}

\begin{thebibliography}{64}
\ifx\bisbn     \undefined \def\bisbn  #1{ISBN #1}\fi
\ifx\binits    \undefined \def\binits#1{#1}\fi
\ifx\bauthor   \undefined \def\bauthor#1{#1}\fi
\ifx\batitle   \undefined \def\batitle#1{#1}\fi
\ifx\bjtitle   \undefined \def\bjtitle#1{\textit{#1}}\fi
\ifx\bvolume   \undefined \def\bvolume#1{\textbf{#1}}\fi
\ifx\byear     \undefined \def\byear#1{#1}\fi
\ifx\bissue    \undefined \def\bissue#1{#1}\fi
\ifx\bfpage    \undefined \def\bfpage#1{#1}\fi
\ifx\blpage    \undefined \def\blpage #1{#1}\fi
\ifx\burl      \undefined \def\burl#1{#1}\fi
\ifx\href      \undefined \def\href#1#2{#2}\fi
\ifx\betal     \undefined \def\betal{et al.}\fi
\ifx\bctitle   \undefined \def\bctitle#1{#1}\fi
\ifx\beditor   \undefined \def\beditor#1{#1}\fi
\ifx\bbtitle   \undefined \def\bbtitle#1{\textit{#1}}\fi
\ifx\bedition  \undefined \def\bedition#1{#1}\fi
\ifx\bseriesno \undefined \def\bseriesno#1{\textbf{#1}}\fi
\ifx\blocation \undefined \def\blocation#1{#1}\fi
\ifx\bsertitle \undefined \def\bsertitle#1{\textit{#1}}\fi
\ifx\bsnm      \undefined \def\bsnm#1{#1}\fi
\ifx\bsuffix   \undefined \def\bsuffix#1{#1}\fi
\ifx\bparticle \undefined \def\bparticle#1{#1}\fi
\ifx\barticle  \undefined \def\barticle#1{}\fi
\ifx\binstitute  \undefined \def\binstitute#1{#1}\fi
\ifx\bpublisher  \undefined \def\bpublisher#1{#1}\fi
\ifx\doiurl    \undefined \def\doiurl#1{\href{#1}{DOI}}\fi
\makeatletter
\def\safeHref#1#2#3{\in@{http}{#2}\ifin@\href{#2}{#3}\else\href{#1#2}{#3}\fi}
\makeatother
\ifx\adsurl    \undefined
  \def\adsurl#1{\safeHref{https://ui.adsabs.harvard.edu/abs/}{#1}{ADS}}\fi
\ifx\arxivurl  \undefined
  \def\arxivurl#1{\safeHref{http://arxiv.org/abs/}{#1}{arXiv}}\fi
\ifx\botherref \undefined \def\botherref#1{}\fi
\ifx\url       \undefined \def\url#1{#1}\fi
\ifx\bchapter  \undefined \def\bchapter#1{}\fi
\ifx\bbook     \undefined \def\bbook#1{}\fi
\ifx\bcomment  \undefined \def\bcomment#1{#1}\fi
\ifx\oauthor   \undefined \def\oauthor#1{#1}\fi
\ifx\citeauthoryear \undefined\def \citeauthoryear#1{#1}\fi
\def\endbibitem {}
\ifx\bconflocation  \undefined \def\bconflocation#1{#1} \fi

\bibitem[\protect\citeauthoryear{{Abbett}, {Fisher}, and
  {Fan}}{2001}]{Abbett+2001}
\begin{barticle}
\bauthor{\bsnm{{Abbett}}, \binits{W.P.}},
\bauthor{\bsnm{{Fisher}}, \binits{G.H.}},
\bauthor{\bsnm{{Fan}}, \binits{Y.}}:
\byear{2001},
\batitle{{The Effects of Rotation on the Evolution of Rising Omega Loops in a
  Stratified Model Convection Zone}}.
\bjtitle{\apj}
\bvolume{546},
\bfpage{1194}.
\doiurl{https://doi.org/10.1086/318320}.
\adsurl{2001ApJ...546.1194A}.
\end{barticle}
\endbibitem

\bibitem[\protect\citeauthoryear{{Alley} and
  {Schunker}}{2023}]{AlleySchunker2023}
\begin{barticle}
\bauthor{\bsnm{{Alley}}, \binits{C.S.}},
\bauthor{\bsnm{{Schunker}}, \binits{H.}}:
\byear{2023},
\batitle{{Evolution of the magnetic field and flows of solar active regions
  with persistent magnetic bipoles before emergence}}.
\bjtitle{\pasa}
\bvolume{40},
\bfpage{e059}.
\doiurl{https://doi.org/10.1017/pasa.2023.52}.
\adsurl{2023PASA...40...59A}.
\end{barticle}
\endbibitem

\bibitem[\protect\citeauthoryear{{Babcock}}{1961}]{Babcock1961}
\begin{barticle}
\bauthor{\bsnm{{Babcock}}, \binits{H.W.}}:
\byear{1961},
\batitle{{The Topology of the Sun's Magnetic Field and the 22-YEAR Cycle.}}
\bjtitle{\apj}
\bvolume{133},
\bfpage{572}.
\doiurl{https://doi.org/10.1086/147060}.
\end{barticle}
\endbibitem

\bibitem[\protect\citeauthoryear{{Berger}}{1984}]{Berger84}
\begin{barticle}
\bauthor{\bsnm{{Berger}}, \binits{M.A.}}:
\byear{1984},
\batitle{{Rigorous new limits on magnetic helicity dissipation in the solar
  corona}}.
\bjtitle{Geophys. Astrophys. Fluid. Dyn.}
\bvolume{30},
\bfpage{79}.
\end{barticle}
\endbibitem

\bibitem[\protect\citeauthoryear{{Birch} et~al.}{2016}]{Birchetal2016}
\begin{barticle}
\bauthor{\bsnm{{Birch}}, \binits{A.C.}},
\bauthor{\bsnm{\textbf{{Schunker}, H.}}},
\bauthor{\bsnm{{Braun}}, \binits{D.C.}},
\bauthor{\bsnm{{Cameron}}, \binits{R.}},
\bauthor{\bsnm{{Gizon}}, \binits{L.}},
\bauthor{\bsnm{{L{\"o}ptien}}, \binits{B.}},
\bauthor{\bsnm{{Rempel}}, \binits{M.}}:
\byear{2016},
\batitle{{A low upper limit on the subsurface rise speed of solar active
  regions}}.
\bjtitle{Science Advances}
\bvolume{2},
\bfpage{e1600557}.
\doiurl{https://doi.org/10.1126/sciadv.1600557}.
\adsurl{2016SciA....2E0557B}.
\end{barticle}
\endbibitem

\bibitem[\protect\citeauthoryear{{Birch} et~al.}{2019}]{Birchetal2019}
\begin{barticle}
\bauthor{\bsnm{{Birch}}, \binits{A.C.}},
\bauthor{\bsnm{{Schunker}}, \binits{H.}},
\bauthor{\bsnm{{Braun}}, \binits{D.C.}},
\bauthor{\bsnm{{Gizon}}, \binits{L.}}:
\byear{2019},
\batitle{{Average surface flows before the formation of solar active regions
  and their relationship to the supergranulation pattern}}.
\bjtitle{Astronomy \& Astrophysics}
\bvolume{628},
\bfpage{A37}.
\doiurl{https://doi.org/10.1051/0004-6361/201935591}.
\adsurl{2019A&A...628A..37B}.
\end{barticle}
\endbibitem

\bibitem[\protect\citeauthoryear{{Biswas}, {Karak}, and
  {Cameron}}{2022}]{Biswas+2022}
\begin{barticle}
\bauthor{\bsnm{{Biswas}}, \binits{A.}},
\bauthor{\bsnm{{Karak}}, \binits{B.B.}},
\bauthor{\bsnm{{Cameron}}, \binits{R.}}:
\byear{2022},
\batitle{{Toroidal Flux Loss due to Flux Emergence Explains why Solar Cycles
  Rise Differently but Decay in a Similar Way}}.
\bjtitle{\prl}
\bvolume{129},
\bfpage{241102}.
\doiurl{https://doi.org/10.1103/PhysRevLett.129.241102}.
\adsurl{2022PhRvL.129x1102B}.
\end{barticle}
\endbibitem

\bibitem[\protect\citeauthoryear{{Bray} and
  {Loughhead}}{1979}]{BrayLoughhead1979}
\begin{bbook}
\bauthor{\bsnm{{Bray}}, \binits{R.J.}},
\bauthor{\bsnm{{Loughhead}}, \binits{R.E.}}:
\byear{1979},
\bbtitle{{Sunspots.}}
\adsurl{1979suns.book.....B}.
\end{bbook}
\endbibitem

\bibitem[\protect\citeauthoryear{{Bumba} and {Suda}}{1984}]{BumbaSuda1984}
\begin{barticle}
\bauthor{\bsnm{{Bumba}}, \binits{V.}},
\bauthor{\bsnm{{Suda}}, \binits{J.}}:
\byear{1984},
\batitle{{Processes observable in the photosphere during the formation of an
  active region. II - Development of a usual active region; growth of a SPOT
  penumbra}}.
\bjtitle{Bulletin of the Astronomical Institutes of Czechoslovakia}
\bvolume{35},
\bfpage{28}.
\adsurl{1984BAICz..35...28B}.
\end{barticle}
\endbibitem

\bibitem[\protect\citeauthoryear{{Caligari}, {Moreno-Insertis}, and
  {Schussler}}{1995}]{Caligari+1995}
\begin{barticle}
\bauthor{\bsnm{{Caligari}}, \binits{P.}},
\bauthor{\bsnm{{Moreno-Insertis}}, \binits{F.}},
\bauthor{\bsnm{{Schussler}}, \binits{M.}}:
\byear{1995},
\batitle{{Emerging Flux Tubes in the Solar Convection Zone. I. Asymmetry, Tilt,
  and Emergence Latitude}}.
\bjtitle{\apj}
\bvolume{441},
\bfpage{886}.
\doiurl{https://doi.org/10.1086/175410}.
\adsurl{1995ApJ...441..886C}.
\end{barticle}
\endbibitem

\bibitem[\protect\citeauthoryear{{Cameron} and
  {Sch{\"u}ssler}}{2015}]{Cameronetal2015}
\begin{barticle}
\bauthor{\bsnm{{Cameron}}, \binits{R.}},
\bauthor{\bsnm{{Sch{\"u}ssler}}, \binits{M.}}:
\byear{2015},
\batitle{{The crucial role of surface magnetic fields for the solar dynamo}}.
\bjtitle{Science}
\bvolume{347},
\bfpage{1333}.
\doiurl{https://doi.org/10.1126/science.1261470}.
\end{barticle}
\endbibitem

\bibitem[\protect\citeauthoryear{{Cameron} and
  {Sch{\"u}ssler}}{2023}]{CameronSchussler2023}
\begin{barticle}
\bauthor{\bsnm{{Cameron}}, \binits{R.H.}},
\bauthor{\bsnm{{Sch{\"u}ssler}}, \binits{M.}}:
\byear{2023},
\batitle{{Observationally Guided Models for the Solar Dynamo and the Role of
  the Surface Field}}.
\bjtitle{\ssr}
\bvolume{219},
\bfpage{60}.
\doiurl{https://doi.org/10.1007/s11214-023-01004-7}.
\adsurl{2023SSRv..219...60C}.
\end{barticle}
\endbibitem

\bibitem[\protect\citeauthoryear{{Cameron}, {Dikpati}, and
  {Brandenburg}}{2017}]{Cameron+2017}
\begin{barticle}
\bauthor{\bsnm{{Cameron}}, \binits{R.H.}},
\bauthor{\bsnm{{Dikpati}}, \binits{M.}},
\bauthor{\bsnm{{Brandenburg}}, \binits{A.}}:
\byear{2017},
\batitle{{The Global Solar Dynamo}}.
\bjtitle{\ssr}
\bvolume{210},
\bfpage{367}.
\doiurl{https://doi.org/10.1007/s11214-015-0230-3}.
\adsurl{2017SSRv..210..367C}.
\end{barticle}
\endbibitem

\bibitem[\protect\citeauthoryear{{Cameron} et~al.}{2013}]{Cameron+2013}
\begin{barticle}
\bauthor{\bsnm{{Cameron}}, \binits{R.H.}},
\bauthor{\bsnm{{Dasi-Espuig}}, \binits{M.}},
\bauthor{\bsnm{{Jiang}}, \binits{J.}},
\bauthor{\bsnm{{I{\c{s}}{\i}k}}, \binits{E.}},
\bauthor{\bsnm{{Schmitt}}, \binits{D.}},
\bauthor{\bsnm{{Sch{\"u}ssler}}, \binits{M.}}:
\byear{2013},
\batitle{{Limits to solar cycle predictability: Cross-equatorial flux plumes}}.
\bjtitle{\aap}
\bvolume{557},
\bfpage{A141}.
\doiurl{https://doi.org/10.1051/0004-6361/201321981}.
\adsurl{2013A&A...557A.141C}.
\end{barticle}
\endbibitem

\bibitem[\protect\citeauthoryear{{Cameron} et~al.}{2018}]{Cameronetal2017}
\begin{barticle}
\bauthor{\bsnm{{Cameron}}, \binits{R.H.}},
\bauthor{\bsnm{{Duvall}}, \binits{T.L.}},
\bauthor{\bsnm{{Sch{\"u}ssler}}, \binits{M.}},
\bauthor{\bsnm{{Schunker}}, \binits{H.}}:
\byear{2018},
\batitle{{Observing and modeling the poloidal and toroidal fields of the solar
  dynamo}}.
\bjtitle{\aap}
\bvolume{609},
\bfpage{A56}.
\doiurl{https://doi.org/10.1051/0004-6361/201731481}.
\end{barticle}
\endbibitem

\bibitem[\protect\citeauthoryear{{Cameron} et~al.}{2025}]{Cameron+2025}
\begin{botherref}
\oauthor{\bsnm{{Cameron}}, \binits{R.H.}},
\oauthor{\bsnm{{Schunker}}, \binits{H.}},
\oauthor{\bsnm{{Roland-Batty}}, \binits{W.}},
\oauthor{\bsnm{{Brun}}, \binits{S.}},
\oauthor{\bsnm{{Strugarek}}, \binits{A.}},
\oauthor{\bsnm{{Finley}}, \binits{A.}},
\oauthor{\bsnm{{Birch}}, \binits{A.}},
\oauthor{\bsnm{{Gizon}}, \binits{L.}}:
2025,
{Closing the solar dynamo loop: poloidal field generation at the surface}.
\textit{accepted \aap}.
\end{botherref}
\endbibitem

\bibitem[\protect\citeauthoryear{Charbonneau}{2020}]{Charbonneau2020}
\begin{barticle}
\bauthor{\bsnm{Charbonneau}, \binits{P.}}:
\byear{2020},
\batitle{Dynamo models of the solar cycle}.
\bjtitle{Living Reviews in Solar Physics}
\bvolume{17},
\bfpage{4}.
\bisbn{1614-4961}.
\doiurl{https://doi.org/10.1007/s41116-020-00025-6}.
\burl{https://doi.org/10.1007/s41116-020-00025-6}.
\end{barticle}
\endbibitem

\bibitem[\protect\citeauthoryear{{Cheung} and {Isobe}}{2014}]{Cheung2014}
\begin{barticle}
\bauthor{\bsnm{{Cheung}}, \binits{M.C.M.}},
\bauthor{\bsnm{{Isobe}}, \binits{H.}}:
\byear{2014},
\batitle{{Flux Emergence (Theory)}}.
\bjtitle{Living Reviews in Solar Physics}
\bvolume{11},
\bfpage{3}.
\doiurl{https://doi.org/10.12942/lrsp-2014-3}.
\adsurl{2014LRSP...11....3C}.
\end{barticle}
\endbibitem

\bibitem[\protect\citeauthoryear{{Dasi-Espuig}
  et~al.}{2010}]{DasiEspuigetal2010}
\begin{barticle}
\bauthor{\bsnm{{Dasi-Espuig}}, \binits{M.}},
\bauthor{\bsnm{{Solanki}}, \binits{S.K.}},
\bauthor{\bsnm{{Krivova}}, \binits{N.A.}},
\bauthor{\bsnm{{Cameron}}, \binits{R.}},
\bauthor{\bsnm{{Pe{\~n}uela}}, \binits{T.}}:
\byear{2010},
\batitle{{Sunspot group tilt angles and the strength of the solar cycle}}.
\bjtitle{\aap}
\bvolume{518},
\bfpage{A7}.
\doiurl{https://doi.org/10.1051/0004-6361/201014301}.
\end{barticle}
\endbibitem

\bibitem[\protect\citeauthoryear{{D'Silva} and
  {Choudhuri}}{1993}]{DSilvaChoudhuri1993}
\begin{barticle}
\bauthor{\bsnm{{D'Silva}}, \binits{S.}},
\bauthor{\bsnm{{Choudhuri}}, \binits{A.R.}}:
\byear{1993},
\batitle{{A theoretical model for tilts of bipolar magnetic regions}}.
\bjtitle{\aap}
\bvolume{272},
\bfpage{621}.
\end{barticle}
\endbibitem

\bibitem[\protect\citeauthoryear{{Durrant}, {Turner}, and
  {Wilson}}{2004}]{Durrant+2004}
\begin{barticle}
\bauthor{\bsnm{{Durrant}}, \binits{C.J.}},
\bauthor{\bsnm{{Turner}}, \binits{J.P.R.}},
\bauthor{\bsnm{{Wilson}}, \binits{P.R.}}:
\byear{2004},
\batitle{{The Mechanism involved in the Reversals of the Sun's Polar Magnetic
  Fields}}.
\bjtitle{\solphys}
\bvolume{222},
\bfpage{345}.
\doiurl{https://doi.org/10.1023/B:SOLA.0000043577.33961.82}.
\adsurl{2004SoPh..222..345D}.
\end{barticle}
\endbibitem

\bibitem[\protect\citeauthoryear{{Fan}}{2008}]{Fan2008}
\begin{barticle}
\bauthor{\bsnm{{Fan}}, \binits{Y.}}:
\byear{2008},
\batitle{{The Three-dimensional Evolution of Buoyant Magnetic Flux Tubes in a
  Model Solar Convective Envelope}}.
\bjtitle{\apj}
\bvolume{676},
\bfpage{680}.
\doiurl{https://doi.org/10.1086/527317}.
\adsurl{2008ApJ...676..680F}.
\end{barticle}
\endbibitem

\bibitem[\protect\citeauthoryear{{Fan}}{2009}]{Fan2009}
\begin{barticle}
\bauthor{\bsnm{{Fan}}, \binits{Y.}}:
\byear{2009},
\batitle{{Magnetic Fields in the Solar Convection Zone}}.
\bjtitle{Living Reviews in Solar Physics}
\bvolume{6},
\bfpage{4}.
\doiurl{https://doi.org/10.12942/lrsp-2009-4}.
\end{barticle}
\endbibitem

\bibitem[\protect\citeauthoryear{{Fan}}{2021}]{Fan2021LRSP}
\begin{barticle}
\bauthor{\bsnm{{Fan}}, \binits{Y.}}:
\byear{2021},
\batitle{{Magnetic fields in the solar convection zone}}.
\bjtitle{Living Reviews in Solar Physics}
\bvolume{18},
\bfpage{5}.
\doiurl{https://doi.org/10.1007/s41116-021-00031-2}.
\adsurl{2021LRSP...18....5F}.
\end{barticle}
\endbibitem

\bibitem[\protect\citeauthoryear{{Fan} and {Fang}}{2014}]{FanFang2014}
\begin{barticle}
\bauthor{\bsnm{{Fan}}, \binits{Y.}},
\bauthor{\bsnm{{Fang}}, \binits{F.}}:
\byear{2014},
\batitle{{A Simulation of Convective Dynamo in the Solar Convective Envelope:
  Maintenance of the Solar-like Differential Rotation and Emerging Flux}}.
\bjtitle{\apj}
\bvolume{789},
\bfpage{35}.
\doiurl{https://doi.org/10.1088/0004-637X/789/1/35}.
\adsurl{2014ApJ...789...35F}.
\end{barticle}
\endbibitem

\bibitem[\protect\citeauthoryear{{Fisher}, {Fan}, and
  {Howard}}{1995}]{Fisher+1995}
\begin{barticle}
\bauthor{\bsnm{{Fisher}}, \binits{G.H.}},
\bauthor{\bsnm{{Fan}}, \binits{Y.}},
\bauthor{\bsnm{{Howard}}, \binits{R.F.}}:
\byear{1995},
\batitle{{Comparisons between theory and observation of active region tilts}}.
\bjtitle{\apj}
\bvolume{438},
\bfpage{463}.
\doiurl{https://doi.org/10.1086/175090}.
\end{barticle}
\endbibitem

\bibitem[\protect\citeauthoryear{Gosain et~al.}{2018}]{Gosain+2018}
\begin{bchapter}
\bauthor{\bsnm{Gosain}, \binits{S.}},
\bauthor{\bsnm{Roth}, \binits{M.}},
\bauthor{\bsnm{Hill}, \binits{F.}},
\bauthor{\bsnm{Pevtsov}, \binits{A.}},
\bauthor{\bsnm{Pillet}, \binits{V.M.}},
\bauthor{\bsnm{Thompson}, \binits{M.J.}}:
\byear{2018},
\bctitle{{Design of a next generation synoptic solar observing network: solar
  physics research integrated network group (SPRING)}}.
In: \beditor{\bsnm{Evans}, \binits{C.J.}},
\beditor{\bsnm{Simard}, \binits{L.}},
\beditor{\bsnm{Takami}, \binits{H.}} (eds.)
\bbtitle{Ground-based and Airborne Instrumentation for Astronomy VII}
\bseriesno{10702},
\bpublisher{SPIE},
\bfpage{107024H}.
\bcomment{International Society for Optics and Photonics}.
\doiurl{https://doi.org/10.1117/12.2306555}.
\burl{https://doi.org/10.1117/12.2306555}.
\end{bchapter}
\endbibitem

\bibitem[\protect\citeauthoryear{{Gottschling} et~al.}{2021}]{Gottschling+2021}
\begin{barticle}
\bauthor{\bsnm{{Gottschling}}, \binits{N.}},
\bauthor{\bsnm{{Schunker}}, \binits{H.}},
\bauthor{\bsnm{{Birch}}, \binits{A.C.}},
\bauthor{\bsnm{{L{\"o}ptien}}, \binits{B.}},
\bauthor{\bsnm{{Gizon}}, \binits{L.}}:
\byear{2021},
\batitle{{Evolution of solar surface inflows around emerging active regions}}.
\bjtitle{\aap}
\bvolume{652},
\bfpage{A148}.
\doiurl{https://doi.org/10.1051/0004-6361/202140324}.
\adsurl{2021A&A...652A.148G}.
\end{barticle}
\endbibitem

\bibitem[\protect\citeauthoryear{{Hale} et~al.}{1919}]{Hale1919}
\begin{barticle}
\bauthor{\bsnm{{Hale}}, \binits{G.E.}},
\bauthor{\bsnm{{Ellerman}}, \binits{F.}},
\bauthor{\bsnm{{Nicholson}}, \binits{S.B.}},
\bauthor{\bsnm{{Joy}}, \binits{A.H.}}:
\byear{1919},
\batitle{{The Magnetic Polarity of Sun-Spots}}.
\bjtitle{The Astrophysical Journal}
\bvolume{49},
\bfpage{153}.
\doiurl{https://doi.org/10.1086/142452}.
\adsurl{1919ApJ....49..153H}.
\end{barticle}
\endbibitem

\bibitem[\protect\citeauthoryear{{Hill} et~al.}{2019}]{Hill+2019}
\begin{bchapter}
\bauthor{\bsnm{{Hill}}, \binits{F.}},
\bauthor{\bsnm{{Hammel}}, \binits{H.}},
\bauthor{\bsnm{{Martinez-Pillet}}, \binits{V.}},
\bauthor{\bsnm{{de Wijn}}, \binits{A.}},
\bauthor{\bsnm{{Gosain}}, \binits{S.}},
\bauthor{\bsnm{{Burkepile}}, \binits{J.}},
\bauthor{\bsnm{{Henney}}, \binits{C.J.}},
\bauthor{\bsnm{{McAteer}}, \binits{J.}},
\bauthor{\bsnm{{Bain}}, \binits{H.M.}},
\bauthor{\bsnm{{Manchester}}, \binits{W.}},
\bauthor{\bsnm{{Lin}}, \binits{H.}},
\bauthor{\bsnm{{Roth}}, \binits{M.}},
\bauthor{\bsnm{{Ichimoto}}, \binits{K.}},
\bauthor{\bsnm{{Suematsu}}, \binits{Y.}}:
\byear{2019},
\bctitle{{ngGONG: The Next Generation GONG - A New Solar Synoptic Observational
  Network}}.
In: \bbtitle{Bulletin of the American Astronomical Society}
\bseriesno{51},
\bfpage{74}.
\adsurl{2019BAAS...51g..74H}.
\end{bchapter}
\endbibitem

\bibitem[\protect\citeauthoryear{{Hotta} and {Iijima}}{2020}]{HottaIijima2020}
\begin{barticle}
\bauthor{\bsnm{{Hotta}}, \binits{H.}},
\bauthor{\bsnm{{Iijima}}, \binits{H.}}:
\byear{2020},
\batitle{{On rising magnetic flux tube and formation of sunspots in a deep
  domain}}.
\bjtitle{\mnras}
\bvolume{494},
\bfpage{2523}.
\doiurl{https://doi.org/10.1093/mnras/staa844}.
\adsurl{2020MNRAS.494.2523H}.
\end{barticle}
\endbibitem

\bibitem[\protect\citeauthoryear{{Kosovichev} and
  {Stenflo}}{2008}]{KosovichevStenflo2008}
\begin{barticle}
\bauthor{\bsnm{{Kosovichev}}, \binits{A.G.}},
\bauthor{\bsnm{{Stenflo}}, \binits{J.O.}}:
\byear{2008},
\batitle{{Tilt of Emerging Bipolar Magnetic Regions on the Sun}}.
\bjtitle{\apjl}
\bvolume{688},
\bfpage{L115}.
\doiurl{https://doi.org/10.1086/595619}.
\end{barticle}
\endbibitem

\bibitem[\protect\citeauthoryear{{Leighton}}{1964}]{Leighton1964}
\begin{barticle}
\bauthor{\bsnm{{Leighton}}, \binits{R.B.}}:
\byear{1964},
\batitle{{Transport of Magnetic Fields on the Sun.}}
\bjtitle{\apj}
\bvolume{140},
\bfpage{1547}.
\doiurl{https://doi.org/10.1086/148058}.
\adsurl{1964ApJ...140.1547L}.
\end{barticle}
\endbibitem

\bibitem[\protect\citeauthoryear{{Leighton}}{1969}]{Leighton1969}
\begin{barticle}
\bauthor{\bsnm{{Leighton}}, \binits{R.B.}}:
\byear{1969},
\batitle{{A Magneto-Kinematic Model of the Solar Cycle}}.
\bjtitle{\apj}
\bvolume{156},
\bfpage{1}.
\doiurl{https://doi.org/10.1086/149943}.
\adsurl{1969ApJ...156....1L}.
\end{barticle}
\endbibitem

\bibitem[\protect\citeauthoryear{{Leka} et~al.}{2013}]{Lekaetal2013}
\begin{barticle}
\bauthor{\bsnm{{Leka}}, \binits{K.D.}},
\bauthor{\bsnm{{Barnes}}, \binits{G.}},
\bauthor{\bsnm{{Birch}}, \binits{A.C.}},
\bauthor{\bsnm{{Gonzalez-Hernandez}}, \binits{I.}},
\bauthor{\bsnm{{Dunn}}, \binits{T.}},
\bauthor{\bsnm{{Javornik}}, \binits{B.}},
\bauthor{\bsnm{{Braun}}, \binits{D.C.}}:
\byear{2013},
\batitle{{Helioseismology of Pre-emerging Active Regions. I. Overview, Data,
  and Target Selection Criteria}}.
\bjtitle{\apj}
\bvolume{762},
\bfpage{130}.
\doiurl{https://doi.org/10.1088/0004-637X/762/2/130}.
\end{barticle}
\endbibitem

\bibitem[\protect\citeauthoryear{{Longcope} and
  {Klapper}}{1997}]{LongcopeKlapper1997}
\begin{barticle}
\bauthor{\bsnm{{Longcope}}, \binits{D.W.}},
\bauthor{\bsnm{{Klapper}}, \binits{I.}}:
\byear{1997},
\batitle{{Dynamics of a Thin Twisted Flux Tube}}.
\bjtitle{\apj}
\bvolume{488},
\bfpage{443}.
\doiurl{https://doi.org/10.1086/304680}.
\adsurl{1997ApJ...488..443L}.
\end{barticle}
\endbibitem

\bibitem[\protect\citeauthoryear{{Longcope}, {Fisher}, and
  {Pevtsov}}{1998}]{Longcope+1998}
\begin{barticle}
\bauthor{\bsnm{{Longcope}}, \binits{D.W.}},
\bauthor{\bsnm{{Fisher}}, \binits{G.H.}},
\bauthor{\bsnm{{Pevtsov}}, \binits{A.A.}}:
\byear{1998},
\batitle{{Flux-Tube Twist Resulting from Helical Turbulence: The
  {\ensuremath{\Sigma}}-Effect}}.
\bjtitle{\apj}
\bvolume{507},
\bfpage{417}.
\doiurl{https://doi.org/10.1086/306312}.
\adsurl{1998ApJ...507..417L}.
\end{barticle}
\endbibitem

\bibitem[\protect\citeauthoryear{{Mani} et~al.}{2024}]{Mani+2024}
\begin{barticle}
\bauthor{\bsnm{{Mani}}, \binits{P.}},
\bauthor{\bsnm{{Hanson}}, \binits{C.S.}},
\bauthor{\bsnm{{Dhanpal}}, \binits{S.}},
\bauthor{\bsnm{{Hanasoge}}, \binits{S.}},
\bauthor{\bsnm{{Das}}, \binits{S.B.}},
\bauthor{\bsnm{{Rempel}}, \binits{M.}}:
\byear{2024},
\batitle{{Magnetic Flux in the Sun Emerges Unaffected by Supergranular-scale
  Surface Flows}}.
\bjtitle{\apj}
\bvolume{965},
\bfpage{186}.
\doiurl{https://doi.org/10.3847/1538-4357/ad2ae3}.
\adsurl{2024ApJ...965..186M}.
\end{barticle}
\endbibitem

\bibitem[\protect\citeauthoryear{{McClintock} and
  {Norton}}{2013}]{McClintockNorton2013}
\begin{barticle}
\bauthor{\bsnm{{McClintock}}, \binits{B.H.}},
\bauthor{\bsnm{{Norton}}, \binits{A.A.}}:
\byear{2013},
\batitle{{Recovering Joy's Law as a Function of Solar Cycle, Hemisphere, and
  Longitude}}.
\bjtitle{\solphys}
\bvolume{287},
\bfpage{215}.
\doiurl{https://doi.org/10.1007/s11207-013-0338-0}.
\adsurl{2013SoPh..287..215M}.
\end{barticle}
\endbibitem

\bibitem[\protect\citeauthoryear{{Mu{\~n}oz-Jaramillo}
  et~al.}{2013}]{MunozJaramillo+2013}
\begin{barticle}
\bauthor{\bsnm{{Mu{\~n}oz-Jaramillo}}, \binits{A.}},
\bauthor{\bsnm{{Dasi-Espuig}}, \binits{M.}},
\bauthor{\bsnm{{Balmaceda}}, \binits{L.A.}},
\bauthor{\bsnm{{DeLuca}}, \binits{E.E.}}:
\byear{2013},
\batitle{{Solar Cycle Propagation, Memory, and Prediction: Insights from a
  Century of Magnetic Proxies}}.
\bjtitle{\apjl}
\bvolume{767},
\bfpage{L25}.
\doiurl{https://doi.org/10.1088/2041-8205/767/2/L25}.
\adsurl{2013ApJ...767L..25M}.
\end{barticle}
\endbibitem

\bibitem[\protect\citeauthoryear{{Nelson} et~al.}{2014}]{Nelson+2014}
\begin{barticle}
\bauthor{\bsnm{{Nelson}}, \binits{N.J.}},
\bauthor{\bsnm{{Brown}}, \binits{B.P.}},
\bauthor{\bsnm{{Sacha Brun}}, \binits{A.}},
\bauthor{\bsnm{{Miesch}}, \binits{M.S.}},
\bauthor{\bsnm{{Toomre}}, \binits{J.}}:
\byear{2014},
\batitle{{Buoyant Magnetic Loops Generated by Global Convective Dynamo
  Action}}.
\bjtitle{\solphys}
\bvolume{289},
\bfpage{441}.
\doiurl{https://doi.org/10.1007/s11207-012-0221-4}.
\end{barticle}
\endbibitem

\bibitem[\protect\citeauthoryear{{Parker}}{1955}]{Parker1955}
\begin{barticle}
\bauthor{\bsnm{{Parker}}, \binits{E.N.}}:
\byear{1955},
\batitle{{The Formation of Sunspots from the Solar Toroidal Field.}}
\bjtitle{\apj}
\bvolume{121},
\bfpage{491}.
\doiurl{https://doi.org/10.1086/146010}.
\adsurl{1955ApJ...121..491P}.
\end{barticle}
\endbibitem

\bibitem[\protect\citeauthoryear{{Pevtsov} et~al.}{2014}]{Pevtsovetal2014}
\begin{barticle}
\bauthor{\bsnm{{Pevtsov}}, \binits{A.A.}},
\bauthor{\bsnm{{Berger}}, \binits{M.A.}},
\bauthor{\bsnm{{Nindos}}, \binits{A.}},
\bauthor{\bsnm{{Norton}}, \binits{A.A.}},
\bauthor{\bsnm{{van Driel-Gesztelyi}}, \binits{L.}}:
\byear{2014},
\batitle{{Magnetic Helicity, Tilt, and Twist}}.
\bjtitle{\ssr}
\bvolume{186},
\bfpage{285}.
\doiurl{https://doi.org/10.1007/s11214-014-0082-2}.
\end{barticle}
\endbibitem

\bibitem[\protect\citeauthoryear{{Rempel} and
  {Cheung}}{2014}]{RempelCheung2014}
\begin{barticle}
\bauthor{\bsnm{{Rempel}}, \binits{M.}},
\bauthor{\bsnm{{Cheung}}, \binits{M.C.M.}}:
\byear{2014},
\batitle{{Numerical Simulations of Active Region Scale Flux Emergence: From
  Spot Formation to Decay}}.
\bjtitle{\apj}
\bvolume{785},
\bfpage{90}.
\doiurl{https://doi.org/10.1088/0004-637X/785/2/90}.
\adsurl{2014ApJ...785...90R}.
\end{barticle}
\endbibitem

\bibitem[\protect\citeauthoryear{{Rincon} and
  {Rieutord}}{2018}]{RinconRieutord2018}
\begin{barticle}
\bauthor{\bsnm{{Rincon}}, \binits{F.}},
\bauthor{\bsnm{{Rieutord}}, \binits{M.}}:
\byear{2018},
\batitle{{The Sun's supergranulation}}.
\bjtitle{Living Reviews in Solar Physics}
\bvolume{15},
\bfpage{6}.
\doiurl{https://doi.org/10.1007/s41116-018-0013-5}.
\adsurl{2018LRSP...15....6R}.
\end{barticle}
\endbibitem

\bibitem[\protect\citeauthoryear{{Roland-Batty}
  et~al.}{2025}]{RolandBatty+2025}
\begin{botherref}
\oauthor{\bsnm{{Roland-Batty}}, \binits{W.}},
\oauthor{\bsnm{{Schunker}}, \binits{H.}},
\oauthor{\bsnm{{Cameron}}, \binits{R.H.}},
\oauthor{\bsnm{{Przybylski}}, \binits{D.}},
\oauthor{\bsnm{{Pontin}}, \binits{D.I.}},
\oauthor{},
\oauthor{\bsnm{{Gizon}}, \binits{L.}}:
2025,
{Coriolis force acting on near-surface horizontal flows during simulations of
  flux emergence produces a tilt angle consistent with Joy’s law on the Sun}.
\textit{\aap}.
\end{botherref}
\endbibitem

\bibitem[\protect\citeauthoryear{{Scherrer} et~al.}{1995}]{SOHO1995}
\begin{barticle}
\bauthor{\bsnm{{Scherrer}}, \binits{P.H.}},
\bauthor{\bsnm{{Bogart}}, \binits{R.S.}},
\bauthor{\bsnm{{Bush}}, \binits{R.I.}},
\bauthor{\bsnm{{Hoeksema}}, \binits{J.T.}},
\bauthor{\bsnm{{Kosovichev}}, \binits{A.G.}},
\bauthor{\bsnm{{Schou}}, \binits{J.}},
\bauthor{\bsnm{{Rosenberg}}, \binits{W.}},
\bauthor{\bsnm{{Springer}}, \binits{L.}},
\bauthor{\bsnm{{Tarbell}}, \binits{T.D.}},
\bauthor{\bsnm{{Title}}, \binits{A.}},
\bauthor{\bsnm{{Wolfson}}, \binits{C.J.}},
\bauthor{\bsnm{{Zayer}}, \binits{I.}},
\bauthor{\bsnm{{MDI Engineering Team}}}:
\byear{1995},
\batitle{{The Solar Oscillations Investigation - Michelson Doppler Imager}}.
\bjtitle{\solphys}
\bvolume{162},
\bfpage{129}.
\doiurl{https://doi.org/10.1007/BF00733429}.
\end{barticle}
\endbibitem

\bibitem[\protect\citeauthoryear{{Scherrer} et~al.}{2012}]{scherreretal2012}
\begin{barticle}
\bauthor{\bsnm{{Scherrer}}, \binits{P.H.}},
\bauthor{\bsnm{{Schou}}, \binits{J.}},
\bauthor{\bsnm{{Bush}}, \binits{R.I.}},
\bauthor{\bsnm{{Kosovichev}}, \binits{A.G.}},
\bauthor{\bsnm{{Bogart}}, \binits{R.S.}},
\bauthor{\bsnm{{Hoeksema}}, \binits{J.T.}},
\bauthor{\bsnm{{Liu}}, \binits{Y.}},
\bauthor{\bsnm{{Duvall}}, \binits{T.L.}},
\bauthor{\bsnm{{Zhao}}, \binits{J.}},
\bauthor{\bsnm{{Title}}, \binits{A.M.}},
\bauthor{\bsnm{{Schrijver}}, \binits{C.J.}},
\bauthor{\bsnm{{Tarbell}}, \binits{T.D.}},
\bauthor{\bsnm{{Tomczyk}}, \binits{S.}}:
\byear{2012},
\batitle{{The Helioseismic and Magnetic Imager (HMI) Investigation for the
  Solar Dynamics Observatory (SDO)}}.
\bjtitle{Solar Physics}
\bvolume{275}.
\doiurl{https://doi.org/10.1007/s11207-011-9834-2}.
\end{barticle}
\endbibitem

\bibitem[\protect\citeauthoryear{{Schlichenmaier}
  et~al.}{2010}]{Schlichenmaier+2010}
\begin{barticle}
\bauthor{\bsnm{{Schlichenmaier}}, \binits{R.}},
\bauthor{\bsnm{{Bello Gonz{\'a}lez}}, \binits{N.}},
\bauthor{\bsnm{{Rezaei}}, \binits{R.}},
\bauthor{\bsnm{{Waldmann}}, \binits{T.A.}}:
\byear{2010},
\batitle{{The role of emerging bipoles in the formation of a sunspot
  penumbra}}.
\bjtitle{Astronomische Nachrichten}
\bvolume{331},
\bfpage{563}.
\doiurl{https://doi.org/10.1002/asna.201011372}.
\adsurl{2010AN....331..563S}.
\end{barticle}
\endbibitem

\bibitem[\protect\citeauthoryear{{Schou} et~al.}{2012}]{Schou+2012}
\begin{barticle}
\bauthor{\bsnm{{Schou}}, \binits{J.}},
\bauthor{\bsnm{{Scherrer}}, \binits{P.H.}},
\bauthor{\bsnm{{Bush}}, \binits{R.I.}},
\bauthor{\bsnm{{Wachter}}, \binits{R.}},
\bauthor{\bsnm{{Couvidat}}, \binits{S.}},
\bauthor{\bsnm{{Rabello-Soares}}, \binits{M.C.}},
\bauthor{\bsnm{{Bogart}}, \binits{R.S.}},
\bauthor{\bsnm{{Hoeksema}}, \binits{J.T.}},
\bauthor{\bsnm{{Liu}}, \binits{Y.}},
\bauthor{\bsnm{{Duvall}}, \binits{T.L.}},
\bauthor{\bsnm{{Akin}}, \binits{D.J.}},
\bauthor{\bsnm{{Allard}}, \binits{B.A.}},
\bauthor{\bsnm{{Miles}}, \binits{J.W.}},
\bauthor{\bsnm{{Rairden}}, \binits{R.}},
\bauthor{\bsnm{{Shine}}, \binits{R.A.}},
\bauthor{\bsnm{{Tarbell}}, \binits{T.D.}},
\bauthor{\bsnm{{Title}}, \binits{A.M.}},
\bauthor{\bsnm{{Wolfson}}, \binits{C.J.}},
\bauthor{\bsnm{{Elmore}}, \binits{D.F.}},
\bauthor{\bsnm{{Norton}}, \binits{A.A.}},
\bauthor{\bsnm{{Tomczyk}}, \binits{S.}}:
\byear{2012},
\batitle{{Design and Ground Calibration of the Helioseismic and Magnetic Imager
  (HMI) Instrument on the Solar Dynamics Observatory (SDO)}}.
\bjtitle{\solphys}
\bvolume{275},
\bfpage{229}.
\doiurl{https://doi.org/10.1007/s11207-011-9842-2}.
\adsurl{2012SoPh..275..229S}.
\end{barticle}
\endbibitem

\bibitem[\protect\citeauthoryear{{Schunker} et~al.}{2016}]{Schunkeretal2016}
\begin{barticle}
\bauthor{\bsnm{{Schunker}}, \binits{H.}},
\bauthor{\bsnm{{Braun}}, \binits{D.C.}},
\bauthor{\bsnm{{Birch}}, \binits{A.C.}},
\bauthor{\bsnm{{Burston}}, \binits{R.B.}},
\bauthor{\bsnm{{Gizon}}, \binits{L.}}:
\byear{2016},
\batitle{{SDO/HMI survey of emerging active regions for helioseismology}}.
\bjtitle{\aap}
\bvolume{595},
\bfpage{A107}.
\doiurl{https://doi.org/10.1051/0004-6361/201628388}.
\adsurl{2016A&A...595A.107S}.
\end{barticle}
\endbibitem

\bibitem[\protect\citeauthoryear{{Schunker} et~al.}{2019}]{Schunkeretal2019}
\begin{barticle}
\bauthor{\bsnm{{Schunker}}, \binits{H.}},
\bauthor{\bsnm{{Birch}}, \binits{A.C.}},
\bauthor{\bsnm{{Cameron}}, \binits{R.H.}},
\bauthor{\bsnm{{Braun}}, \binits{D.C.}},
\bauthor{\bsnm{{Gizon}}, \binits{L.}},
\bauthor{\bsnm{{Burston}}, \binits{R.B.}}:
\byear{2019},
\batitle{{Average motion of emerging solar active region polarities. I. Two
  phases of emergence}}.
\bjtitle{\aap}
\bvolume{625},
\bfpage{A53}.
\doiurl{https://doi.org/10.1051/0004-6361/201834627}.
\adsurl{2019A&A...625A..53S}.
\end{barticle}
\endbibitem

\bibitem[\protect\citeauthoryear{{Schunker} et~al.}{2020}]{Schunkeretal2020}
\begin{barticle}
\bauthor{\bsnm{{Schunker}}, \binits{H.}},
\bauthor{\bsnm{{Baumgartner}}, \binits{C.}},
\bauthor{\bsnm{{Birch}}, \binits{A.C.}},
\bauthor{\bsnm{{Cameron}}, \binits{R.H.}},
\bauthor{\bsnm{{Braun}}, \binits{D.C.}},
\bauthor{\bsnm{{Gizon}}, \binits{L.}}:
\byear{2020},
\batitle{{Average motion of emerging solar active region polarities. II. Joy's
  law}}.
\bjtitle{\aap}
\bvolume{640},
\bfpage{A116}.
\doiurl{https://doi.org/10.1051/0004-6361/201937322}.
\adsurl{2020A&A...640A.116S}.
\end{barticle}
\endbibitem

\bibitem[\protect\citeauthoryear{{Schunker} et~al.}{2024}]{Schunker+2024}
\begin{botherref}
\oauthor{\bsnm{{Schunker}}, \binits{H.}},
\oauthor{\bsnm{{Roland-Batty}}, \binits{W.}},
\oauthor{\bsnm{{Birch}}, \binits{A.C.}},
\oauthor{\bsnm{{Braun}}, \binits{D.C.}},
\oauthor{\bsnm{{Cameron}}, \binits{R.H.}},
\oauthor{\bsnm{{Gizon}}, \binits{L.}}:
2024,
{A flux-independent increase in outflows prior to the emergence of active
  regions on the Sun}.
\textit{\mnras}.
\end{botherref}
\endbibitem

\bibitem[\protect\citeauthoryear{{Sreedevi} et~al.}{2024}]{Sreedevi+2024}
\begin{barticle}
\bauthor{\bsnm{{Sreedevi}}, \binits{A.}},
\bauthor{\bsnm{{Jha}}, \binits{B.K.}},
\bauthor{\bsnm{{Karak}}, \binits{B.B.}},
\bauthor{\bsnm{{Banerjee}}, \binits{D.}}:
\byear{2024},
\batitle{{Analysis of BMR Tilt from AutoTAB Catalog: Hinting toward the Thin
  Flux Tube Model?}}
\bjtitle{\apj}
\bvolume{966},
\bfpage{112}.
\doiurl{https://doi.org/10.3847/1538-4357/ad34b8}.
\adsurl{2024ApJ...966..112S}.
\end{barticle}
\endbibitem

\bibitem[\protect\citeauthoryear{{Steenbeck} and
  {Krause}}{1969}]{SteenbeckKrause1969}
\begin{barticle}
\bauthor{\bsnm{{Steenbeck}}, \binits{M.}},
\bauthor{\bsnm{{Krause}}, \binits{F.}}:
\byear{1969},
\batitle{{On the Dynamo Theory of Stellar and Planetary Magnetic Fields. I. AC
  Dynamos of Solar Type}}.
\bjtitle{Astronomische Nachrichten}
\bvolume{291},
\bfpage{49}.
\doiurl{https://doi.org/10.1002/asna.19692910201}.
\adsurl{1969AN....291...49S}.
\end{barticle}
\endbibitem

\bibitem[\protect\citeauthoryear{{Stenflo} and
  {Kosovichev}}{2012}]{StenfloKosovichev2012}
\begin{barticle}
\bauthor{\bsnm{{Stenflo}}, \binits{J.O.}},
\bauthor{\bsnm{{Kosovichev}}, \binits{A.G.}}:
\byear{2012},
\batitle{{Bipolar Magnetic Regions on the Sun: Global Analysis of the SOHO/MDI
  Data Set}}.
\bjtitle{\apj}
\bvolume{745},
\bfpage{129}.
\doiurl{https://doi.org/10.1088/0004-637X/745/2/129}.
\end{barticle}
\endbibitem

\bibitem[\protect\citeauthoryear{{Talafha} et~al.}{2022}]{Talafha+2022}
\begin{barticle}
\bauthor{\bsnm{{Talafha}}, \binits{M.}},
\bauthor{\bsnm{{Nagy}}, \binits{M.}},
\bauthor{\bsnm{{Lemerle}}, \binits{A.}},
\bauthor{\bsnm{{Petrovay}}, \binits{K.}}:
\byear{2022},
\batitle{{Role of observable nonlinearities in solar cycle modulation}}.
\bjtitle{\aap}
\bvolume{660},
\bfpage{A92}.
\doiurl{https://doi.org/10.1051/0004-6361/202142572}.
\adsurl{2022A&A...660A..92T}.
\end{barticle}
\endbibitem

\bibitem[\protect\citeauthoryear{{Wang} and {Sheeley}}{1991}]{WangSheeley1991}
\begin{barticle}
\bauthor{\bsnm{{Wang}}, \binits{Y.-M.}},
\bauthor{\bsnm{{Sheeley}}, \binits{N.R.} \bsuffix{Jr.}}:
\byear{1991},
\batitle{{Magnetic flux transport and the sun's dipole moment - New twists to
  the Babcock-Leighton model}}.
\bjtitle{\apj}
\bvolume{375},
\bfpage{761}.
\doiurl{https://doi.org/10.1086/170240}.
\end{barticle}
\endbibitem

\bibitem[\protect\citeauthoryear{{Wang} et~al.}{2015}]{Wang+2015}
\begin{barticle}
\bauthor{\bsnm{{Wang}}, \binits{Y.-M.}},
\bauthor{\bsnm{{Colaninno}}, \binits{R.C.}},
\bauthor{\bsnm{{Baranyi}}, \binits{T.}},
\bauthor{\bsnm{{Li}}, \binits{J.}}:
\byear{2015},
\batitle{{Active-region Tilt Angles: Magnetic versus White-light Determinations
  of Joy's Law}}.
\bjtitle{\apj}
\bvolume{798},
\bfpage{50}.
\doiurl{https://doi.org/10.1088/0004-637X/798/1/50}.
\adsurl{2015ApJ...798...50W}.
\end{barticle}
\endbibitem

\bibitem[\protect\citeauthoryear{{Weber}, {Fan}, and
  {Miesch}}{2011}]{Weberetal2011}
\begin{barticle}
\bauthor{\bsnm{{Weber}}, \binits{M.A.}},
\bauthor{\bsnm{{Fan}}, \binits{Y.}},
\bauthor{\bsnm{{Miesch}}, \binits{M.S.}}:
\byear{2011},
\batitle{{The Rise of Active Region Flux Tubes in the Turbulent Solar
  Convective Envelope}}.
\bjtitle{\apj}
\bvolume{741},
\bfpage{11}.
\doiurl{https://doi.org/10.1088/0004-637X/741/1/11}.
\end{barticle}
\endbibitem

\bibitem[\protect\citeauthoryear{{Weber} et~al.}{2023}]{Weberetal2023}
\begin{barticle}
\bauthor{\bsnm{{Weber}}, \binits{M.A.}},
\bauthor{\bsnm{{Schunker}}, \binits{H.}},
\bauthor{\bsnm{{Jouve}}, \binits{L.}},
\bauthor{\bsnm{{I{\c{s}}{\i}k}}, \binits{E.}}:
\byear{2023},
\batitle{{Understanding Active Region Origins and Emergence on the Sun and
  Other Cool Stars}}.
\bjtitle{\ssr}
\bvolume{219},
\bfpage{63}.
\doiurl{https://doi.org/10.1007/s11214-023-01006-5}.
\adsurl{2023SSRv..219...63W}.
\end{barticle}
\endbibitem

\bibitem[\protect\citeauthoryear{{Will}, {Norton}, and
  {Hoeksema}}{2024}]{Will+2024}
\begin{barticle}
\bauthor{\bsnm{{Will}}, \binits{L.W.}},
\bauthor{\bsnm{{Norton}}, \binits{A.A.}},
\bauthor{\bsnm{{Hoeksema}}, \binits{J.T.}}:
\byear{2024},
\batitle{{The Dependence of Joy's Law and Mean Tilt as a Function of Flux
  Emergence Phase}}.
\bjtitle{\apj}
\bvolume{976},
\bfpage{20}.
\doiurl{https://doi.org/10.3847/1538-4357/ad82e3}.
\adsurl{2024ApJ...976...20W}.
\end{barticle}
\endbibitem

\bibitem[\protect\citeauthoryear{{Yeates} et~al.}{2023}]{Yeates+2023}
\begin{barticle}
\bauthor{\bsnm{{Yeates}}, \binits{A.R.}},
\bauthor{\bsnm{{Cheung}}, \binits{M.C.M.}},
\bauthor{\bsnm{{Jiang}}, \binits{J.}},
\bauthor{\bsnm{{Petrovay}}, \binits{K.}},
\bauthor{\bsnm{{Wang}}, \binits{Y.-M.}}:
\byear{2023},
\batitle{{Surface Flux Transport on the Sun}}.
\bjtitle{\ssr}
\bvolume{219},
\bfpage{31}.
\doiurl{https://doi.org/10.1007/s11214-023-00978-8}.
\adsurl{2023SSRv..219...31Y}.
\end{barticle}
\endbibitem

\end{thebibliography}

\end{document}